\documentclass[twocolumn,prl,aps,epsfig]{revtex4}
\usepackage{epsfig}
\def\beq{\begin{equation}}
\def\eeq{\end{equation}}
\def\beqa{\begin{eqnarray}}
\def\eeqa{\end{eqnarray}}
\def\MeV{\nobreak\,\mbox{MeV}}
\def\GeV{\nobreak\,\mbox{GeV}}

\def\la{\lambda}

\def\lb{\label}

\def\me#1{\langle{#1}\rangle}
\def\mer#1{{\langle{#1}\rangle}_{\rho_N}}

\def\bra#1{\langle #1|}
\def\ket#1{| #1\rangle}
\def\qbar{\overline{q}}
\def\gs{g_{\rm s}}
\def\G{{\cal G}}
\def\mixbar{\gs\qbar\sigma\!\cdot\!\G q}
\def\mixs{\gs\bar{s}\sigma\!\cdot\!\G s}
\def\mixda{\gs q^\dagger\sigma\!\cdot\!\G q}
\def\mixdas{\gs s^\dagger\sigma\!\cdot\!\G s}

\def\gluoncon{{\displaystyle{\gs^2G^2}}}
\def\qslash{\rlap{/}{q}}
\def\uslash{\rlap{/}{u}}

\def\qsq{q^2}
\def\qdu{q\cdot u}
\def\veq{|\vec{q}|}
\def\mte{m_\Theta}
\def\mtes{m_\Theta^*}
\def\late{{\la_\Theta^*}^2}
\def\Eq{E_q^*}
\def\Eqb{\bar{E_q}}

\begin{document}

\title{\sc  A QCD Sum Rule Study of $\Theta^+$ in Nuclear Matter}
\author{F.S. Navarra$^a$, M. Nielsen$^a$ and K. Tsushima$^{b,c}$}
\affiliation{$^a$ Instituto de F\'{\i}sica, Universidade de S\~{a}o Paulo,
 C.P. 66318,  05315-970, S\~{a}o Paulo, SP, Brazil\\
$^b$ IFT - UNESP, Rua Pamplona 145, 01405-900,
            S\~ao Paulo, SP, Brazil\\
$^c$ NFC - FCBEE - Universidade Mackenzie,
            Rua da Consola\c{c}\~ao 930, 01302-907,
            S\~ao Paulo, SP, Brazil}

\begin{abstract}
We consider  a $[ud]^2\bar{s}$ current, in the finite-density QCD sum rule
approach, to investigate the scalar and vector self-energies of the recently 
observed pentaquark state $\Theta^+(1540)$, propagating  in nuclear matter. 
We find that, opposite to what was obtained for the nucleon, the vector
self-energy is negative, and the scalar self-energy is positive. There is a
substantial cancellation between them resulting in an attractive net 
self-energy of the same order as in the nucleon case.
\end{abstract}

\maketitle

\vspace{1cm}

The possible existence of a narrow exotic baryon $\Theta^+$ with 
strangeness +1 
is, nowadays, possibly one of the most exciting topics in nuclear physics. 
There is a large number of experiments with evidence for the existence of this
state \cite{exp}, and a similar number of high-energy experiments that see
no evidence \cite{notheta}. Even lattice QCD simulations of pentaquarks
by several groups have not converged yet \cite{lattice,sasa}.
The apparent contradiction between evidences for 
and against the existence  of $\Theta^+$ might be resolved if there is
a production mechanism which is present in some experiments and absent in
others. One such production mechanism was proposed in refs.~\cite{kali,azim}
and is related with the formation of the cryptoexotic $N^*(2400)$ resonance, 
which would decay into $\Theta^+~K^-$. The $N^*(2400)$ could be formed in
the reactions $\gamma+p\rightarrow\pi^++N^*(2400)$ and $\pi^-+p\rightarrow
N^*(2400)$ \cite{azim}, which are not present in $e^+e^-$ collisions
and could be dificult to happen at high-energy collisions. Other possible
production mechanism is meson exchange between the nucleon and another hadron.
Since at high energies all exchanges with the non-vaccum quantum numbers die
out and only the gluonic pomerons survive, it may be difficult to excite the 
pentaquark by soft gluons on nucleon. \cite{diak}.

If $\Theta^+$ is indeed produced through the decay of the $N^*(2400)$
resonance or through meson exchange,
then one should expect that it could be seen in heavy ion collisions
at RHIC, since these two processes
could happen in the collision of nucleons and 
many ``comoving'' hadrons produced in  the  collisions at RHIC. However, if
this is the case, the  $\Theta^+$ would be formed in a nuclear medium
which could change its mass and decay width. Therefore, to be able to
identify positively the  $\Theta^+$ signal in heavy ion collision at RHIC,
it would be very important to know how the nuclear medium affects the
 $\Theta^+$ characteristics. In this work we will use the
QCD sum rule (QCDSR) approach in nuclear matter \cite{srnm1,srnm2,srnm3} to 
study the pentaquark  $\Theta^+$ in a finite density medium.

The finite density QCDSR approach focuses on a correlation function 
evaluated in the ground
state of nuclear matter, instead of the QCD vaccum (as in the usual sum rules).
For spin-${1\over2}$ baryons, this function can be decomposed into three
invariant functions of two kinematic invariants. The appearence of an 
additional invariant function, compared with the vaccum case, is due to an 
additional four-vector: the four-velocity of the nuclear medium, $u_\mu$,
which, together with the the nuclear density, $\rho_N$, characterizes
the nuclear matter ground state. The quasibaryon excitations  are characterized
by scalar and vector self-energies. By introducing a simple ansatz for the
spectral densities, one obtains a phenomenological representation of the
correlation function.

	The correlation function can be also evaluated at large spacelike
momenta using an operator product expansion (OPE). This expansion is written 
in terms of  the matrix elements of composite quark and gluon 
operators, evaluated in the nuclear matter ground state: the in-medium 
condensates. By equating the OPE and phenomenological representations of
the correlation function, one obtains QCD sum rules that relate the baryon 
self-energies to the in-medium condensates.

Several zero-density QCDSR investigations of $\Theta^+$ already exist
where the authors have used different interpolating fields 
\cite{zhu,math,oka,eide,iof}. Here we use the interpolating field suggested
in refs.~\cite{oka,sasa}:
%
\beq
\eta(x)=\epsilon_{abc}\epsilon_{def}\epsilon_{cfg}[u_a^T(x) C d_b(x)]
[u_d^T(x) C \gamma_5 d_e(x)]C\bar{s}^T_g(x),
\label{eta}
\eeq
%
where $a,~b,~c,...$ are color index and $C=-C^T$ is the charge 
conjugation operator. In Eq.~(\ref{eta}) each diquark
pair has spin and isospin zero and is in the $\bar{\textbf{3}}$ 
representation of color SU(3). The total current has isospin zero, positive 
parity and spin 1/2. In ref.~\cite{oka} it was shown that the ground state
described by this current has negative parity and a mass compatible with
the experimental $\Theta^+$ mass.

The QCDSR for $\Theta^+$ at finite density is based on the correlation function
defined by
\beq
\Pi(q)\equiv i\int d^4 x\, e^{iq\cdot x}
\bra{\Psi_0} T\eta(x)\overline{\eta}(0)\ket{\Psi_0}\ ,
\label{cor}
\eeq
where $\ket{\Psi_0}$ represents the nuclear matter ground state.

The correlator in Eq.~(\ref{cor}) can be decomposed in three distinct 
structures \cite{srnm1}
\beq
\Pi(q)\equiv \Pi_s(\qsq,\qdu)+ \Pi_q(\qsq,\qdu)\qslash+\Pi_u(\qsq,\qdu)\uslash
\,.\lb{stru}
\eeq
In vaccum $\Pi_s$ and $\Pi_q$ become function of $\qsq$ only and $\Pi_u$ 
vanishes. For simplicity we will work in the rest frame of nuclear matter,
which implies that $u_\mu=(1,0)$. Therefore, $\Pi_i(\qsq,\qdu)\rightarrow
\Pi_i(q_0,\veq)\;(i=\{s,q,u\})$.

In the phenomenological side, the analytic properties of $\Pi(q)$ can be 
studied 
through a Lehman representation, which leads to a dispersion relation in
$q_0$, for each invariant function, of the form \cite{srnm1}
\beq
\Pi_i(q_0,\veq)={1\over2\pi i}\int_{-\infty}^\infty~d\omega{\Delta\Pi_i(
\omega,\veq)\over\omega-q_0}\;,
\lb{leh}
\eeq
where we have omitted polynomials arising from the contour at large $|q_0|$,
which will be eliminated by the Borel transformation. The discontinuity,
defined by $\Delta\Pi_i(\omega,\veq)$,
contains the spectral information on the quasiparticle, quasihole, and 
higher-energy states.

In vaccum, the spectral weights for baryon and antibaryon are related by
charge conjugation symmetry and one usually parametrizes the spectral
density as a single sharp pole, representing the lowest resonance, plus a 
smooth continuum, representing higher-mass states. At finite density, the 
ground state is no longer invariant under charge conjugation and, therefore,
the spectral densities for baryon and antibaryon are not simply related.

The width of $\Theta^+$ in free space is very small and can be ignored
on hadronic scales \cite{gas}. At finite density, the width of $\Theta^+$ 
can be
broadened due to strong interactions. Since the introduction of one extra
parameter in the spectral density would reduce the predictive power of the 
sum rule, here we assume that a sharp pole hypotesis is still valid at finite
density. In the context of relativistic phenomenology, we assume that 
$\Theta^+$ couples to the same scalar and vector fields as the nucleon
and the hyperons in nuclear matter. Therefore, we parametize the discontinuites
as \cite{srnm1}
\beqa
\Delta\Pi_s(\omega,\veq)&=&-2\pi i{\pm\late\mtes\over2\Eq}\left[
\delta(\omega-E_q)-\delta(\omega-\Eqb)\right]\,,
\nonumber\\
\Delta\Pi_q(\omega,\veq)&=&-2\pi i{\late\over2\Eq}\left[\delta(\omega-E_q)
-\delta(\omega-\Eqb)\right]\,,
\nonumber\\
\Delta\Pi_u(\omega,\veq)&=&2\pi i{\late\Sigma_u\over2\Eq}\left[\delta
(\omega-E_q)
-\delta(\omega-\Eqb)\right]\,,
\lb{dis}
\eeqa
where $\mtes=m_\Theta+\Sigma_s$, $\Eq=\sqrt{{\mtes}^2+\vec{q}^2}$,
$E_q=\Sigma_v+\Eq$ and $\Eqb=\Sigma_v-\Eq$. $\late$ measures the coupling
of the interpolating field with the physical $\Theta^+$ in the medium and
has opposite signs in $\Delta\Pi_s$ depending on the parity of $\Theta^+$
\cite{klee}.
The scalar and vector self-energies of $\Theta^+$ in nuclear matter
are given by $\Sigma_s$ and $\Sigma_v$ respectively. The positive- and 
negative-energy poles are at $E_q$ and $\Eqb$ respectively. In Eq.~(\ref{dis})
we have omitted the contributions from higher-energy states, which will be
included later in the usual form, {\it i.e.}, being approximated  by the OPE 
spectral density starting at an effective threshold. 

At finite density, the OPE for the invariant functions takes the general form
\cite{srnm1,srnm2}
\beq
\Pi_i(q_0,\veq)=\sum_n C_n^i(q_0,\veq)\me{\hat{O}_n}_{\rho_N}\,,
\eeq
where $\me{\hat{O}_n}_{\rho_N}=\bra{\Psi_0}\hat{O}_n\ket{\Psi_0}$, are the 
in-medium condensates. The Wilson coefficients, $C_n^i(q_0,\veq)$ depend only
on QCD Lagrangian parameters. Therefore, all the density dependence of
the correlator is included in the in-medium condensates. 

Following ref.~\cite{srnm1} we separate the invariant functions into two pieces
that are even and odd in $q_0$:
\beq
\Pi_i(q_0,\veq)=\Pi_i^E(q_0^2,\veq)+q_0\Pi_i^O(q_0^2,\veq)\,.
\eeq
In ref.~\cite{srnm1} it was shown that the choice $q_0=\Eqb$ completely
suppresses sharp excitations at $\Eqb$ and also strongly suppresses a broad
excitation in this vicinity. Since we are interested in the positive
energy pole, we will  use $q_0=\Eqb$.

By using the quark 
propagators given in refs.~\cite{srnm1,srnm2,srnm3} and working to leading 
order in 
perturbation theory, up to dimension 5, we get (after Borel transforming
both sides of the sum rules):
\begin{widetext}
\beqa
\late e^{-(E_q^2-\vec{q}^2)/M^2} &=&{M^{12}E_5\over 2^{10}\pi^85!7}
+{m_s\mer{\bar{s}s}\over2^{8}\pi^65!}M^{8}E_3-{M^8E_3\over2^6\pi^65!3}\left(
\mer{s^\dagger iD_0s}-{1\over4}m_s\mer{\bar{s}s}\right)
+{\mer{\gluoncon}\over2^{12}\pi^85!}M^{8}E_3
\nonumber\\
& - &
{2\mer{q^\dagger iD_0q}\over\pi^64!6!}(2M^8E_3+\vec{q}^2M^6E_2)-{\Eqb
\mer{q^\dagger q}\over2^3\pi^64!5!}M^8E_3+{\Eqb M^6E_2\over2^4\pi^63!5!}
\left(\mer{s^\dagger iD_0iD_0s}\right.
\nonumber\\
&+&\left.{1\over12}\mer{\mixdas}\right)-
{\Eqb\mer{\mixdas}\over2^6\pi^63!5!3}
M^6E_2-{\Eqb (M^6E_2+2\vec{q}^2M^4E_1)\over2^2\pi^64!5!}
\left(\mer{q^\dagger iD_0iD_0q}\right.
\nonumber\\
&+&\left.{1\over12}\mer{\mixda}\right)+{\Eqb\mer{\mixda}\over2^3\pi^64!5!}M^6
E_2\,,
\lb{piq}
\eeqa
\beqa
\pm\late\mtes e^{-(E_q^2-\vec{q}^2)/M^2} &=&{m_sM^{12}E_5\over 2^{10}\pi^85!}
-{\mer{\bar{s}s}\over2^{7}\pi^65!}M^{10}E_4-{m_s\mer{q^\dagger iD_0q}\over2^3
\pi^63!5!}\left(9M^8E_3+4\vec{q}^2M^6E_2\right)
\nonumber\\
&-&{9M^8E_3+4\vec{q}^2M^6E_2\over2^5\pi^63!5!}
\left(\mer{\bar{s}iD_0iD_0s}+{1\over8}\mer{\mixs}\right)
+{\mer{\mixs}\over2^{9}\pi^64!}M^8E_3
\nonumber\\
& + &
{m_s\mer{\gluoncon}\over2^{12}\pi^84!}M^{8}E_3
+{\Eqb m_s M^8E_3\over2^5\pi^65!}
\left(\mer{q^\dagger q}+{\mer{s^\dagger s}\over2}\right)
\lb{pim}
\eeqa
\beqa
\late\Sigma_v e^{-(E_q^2-\vec{q}^2)/M^2} &=&-{M^{10}E_4\over 2^{7}\pi^65!3}
\left(\mer{q^\dagger q}+3\mer{s^\dagger s}\right)
-{(5M^8E_3+2\vec{q}^2M^6E_2)\over2^{6}\pi^65!}\left(\mer{s^\dagger iD_0iD_0 s}
\right.
\nonumber\\
&+&\left.{1\over12}\mer{\mixdas}\right)+
{\mer{\mixdas}\over2^5\pi^65!3}M^8E_3-{(5M^8E_3+2\vec{q}^2M^6E_2)\over2^{6}
\pi^65!}\left(\mer{q^\dagger iD_0iD_0 q}\right.
\nonumber\\
&+&\left.{1\over12}\mer{\mixda}\right)+{\mer{\mixda}\over2^4\pi^64!5!}M^8E_3
-{\Eqb\mer{q^\dagger iD_0 q}\over3\pi^64!5!}M^8E_3
\nonumber\\
&+&{\Eqb M^8E_3\over2^4\pi^65!3}\left(\mer{s^\dagger iD_0 s}-{1\over4}
\mer{q^\dagger iD_0 q}\right)\,,
\lb{piu}
\eeqa
\end{widetext}
where we have defined
\beq
E_n\equiv 1-e^{-s_0/M^2}\sum_{k=0}^n\left(s_0\over M^2\right)^k{1\over k!}
\ ,
\label{con}
\eeq
which accounts for the continuum contribution with $s_0$ being the continuum
threshold.

To extract the self-energies from the above finite density sum rules, one
has to know the values of the in-medium condensates. To first order in the
nucleon density, one can write $\mer{\hat{O}}\sim\me{\hat{O}}+
\me{\hat{O}}_N\rho_N$ where $\me{\hat{O}}_N$ is the spin-averaged nucleon 
matrix element. The simplest in-medium condensates are $\mer{q^\dagger q}$
and $\mer{s^\dagger s}$. Since the baryon current is conserved, $\mer{
q^\dagger q}$ and $\mer{s^\dagger s}$ are proportional to the nucleon and 
strangeness densities:
$\mer{q^\dagger q}={3\over2}\rho_N$ and $\mer{s^\dagger s}=0$. These are exact 
results. We extract the values of the other condensates from 
refs.~\cite{srnm1,srnm2,srnm3}: $\mer{\qbar q}=\me{\qbar q}+{\sigma_N\over 
m_u+m_d}\rho_N $, $\mer{\bar{s} s}=\me{\bar{s} s}+y{\sigma_N\over m_u+m_d}
\rho_N$,
$\mer{\gluoncon}=\me{\gluoncon}-4\pi^2(0.65\GeV)\rho_N$,
$\mer{q^\dagger iD_0 q}=(0.18\GeV)\rho_N$,
$\mer{s^\dagger iD_0 s}-{m_s\over4}\mer{\bar{s} s}=(0.018\GeV)\rho_N$,
$\mer{\mixbar}=m_0^2\mer{\qbar q},\mer{\mixs}=m_0^2\mer{\bar{s} s}$,
$\mer{\qbar iD_0iD_0 q}+{1\over8}\mer{\mixbar}=(0.3\GeV^2)\rho_N$,
$\mer{\bar{s} iD_0iD_0 s}+{1\over8}\mer{\mixs}=y(0.3\GeV^2)\rho_N$,
$\mer{\mixda}=(-0.33\GeV^2)\rho_N,\;\;\mer{\mixdas}=y\mer{\mixda}$,
$\mer{q^\dagger iD_0iD_0 q}+{1\over12}\mer{\mixda}=(0.031\GeV^2)\rho_N$,
$\mer{s^\dagger iD_0iD_0 s}+{1\over12}\mer{\mixdas}=y(0.031\GeV^2)\rho_N$
where $\sigma_N$ is the nucleon $\sigma$ term and $y=\me{\bar{s}s}_N/
\me{\qbar q}_N$ is a parameter that measures the strangeness content of the 
nucleon. In this work we use $\sigma_N=45\MeV$, $m_u+m_d=12\MeV$
 and $y$ in the range $0\leq y\leq0.5$ \cite{srnm3}.
The values used for the strange quark mass and vacuum condensates are the 
same as used in ref.~\cite{oka}: $m_s=0.13\,\GeV$, $\me{\qbar q}=\,
-(0.23)^3\,\GeV^3$,
$\langle\overline{s}s\rangle\,=0.8\me{\qbar q}$, $m_0^2=0.8\,\GeV^2$ and 
$\me{\gluoncon}=4\pi^2(0.33~\GeV)^4$. Nuclear matter saturation density is
taken to be $\rho_N=\rho_0=(110\MeV)^3$. For the value of the three-momentum
we follow refs.\cite{srnm1,srnm2,srnm3} and fix it at $\veq=270\MeV$.

We evaluate our sum rules in 
the range $1.7\leq M^2\leq2.5\GeV^2$. As in ref.~\cite{oka} we found that 
the left-hand side of Eq.(\ref{pim}) is negative indicating that the parity
of $\Theta^+$ is negative \cite{klee}. Since the sum rule in  Eq.(\ref{pim})
is very unstable, in this work we neglect it. To eliminate the
coupling $\late$, we follow refs.~\cite{zhu,math} and derive another sum 
rule by taking the derivative of Eq.(\ref{piq}) with respect to $M^{-2}$ and 
dividing it by Eq.~(\ref{piq}). We also divide Eq.(\ref{piu}) by Eq.(\ref{piq})
obtaining two sum rules that are independent of $\late$. To quantify the fit of
the left- and right-hand sides of the sum rules and to determine the 
self-energies, we apply the logarithmic measure
defined in refs.~\cite{srnm1,srnm2,srnm3}. To check the validity of this 
procedure we start by analyzing the sum rule in vacuum. 
The optimized values for the $\Theta^+$ mass and continuum threshold are
$\mte=1.64\GeV$ and $s_0=3.55\GeV^2$, which are in a good agreement 
with the values found in ref.\cite{oka}.

\begin{figure}[h] \label{fig1}
\centerline{\psfig{figure=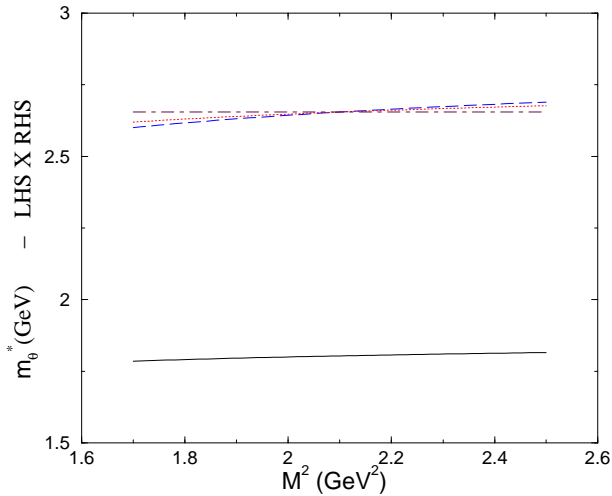,width=8cm,angle=0}}
\caption{The left- and right-sides of the sum rules (upper lines) and the
effective $\Theta^+$ mass (solid line) as functions of the Borel parameter
 $M^2$, for $y=0.3$.}
\end{figure}

In Fig.~1, using $y=0.3$ \cite{str}, we show, in the upper part, 
the left- (dotted line) 
and right-hand sides (dashed and dash-dotted lines) of the sum rules 
(in units of $\GeV^2$) 
as functions of $M^2$. We see that there is a good agreement between them. 
In the lower part of Fig.~1 (solid line) we show the effective $\Theta^+$ 
mass. The optimized values for the effective $\Theta^+$ mass, the vector 
self-energy and continuum threshold are: $\mtes=1.75\GeV$, $\Sigma_v=-150\MeV$
and $s_0=3.6\GeV^2$. The first astonishing result is that, opposite to
what was obtained in the nucleon and hyperon cases, the scalar self-energy
is positive and the vector self-energy is negative. Therefore the effective
$\Theta^+$ mass in medium, $\mtes=\mte+\Sigma_s$, is bigger than
the free $\Theta^+$ mass. However, since $\Sigma_s\simeq110\MeV$, there is
still a substantial  cancelation between $\Sigma_s$ and $\Sigma_v$ in medium.
Looking at Eq.~(\ref{piu}) it is easy to understand why we get $\Sigma_v<0$,
since, in contrast to the nucleon and hyperon cases, all the
terms in the right-hand side are negative. The origin of this difference 
in the sign can 
be attributed to the existence of an antiquark in the $\Theta^+$
interpolating field.
Since $\Sigma_s$ and $\Sigma_v$ are essentially the real parts of the optical 
potential, this qualitative result shows that there is an overall attractive 
$\Theta-$nucleon interaction.

\begin{figure}[h] \label{fig2}
\centerline{\psfig{figure=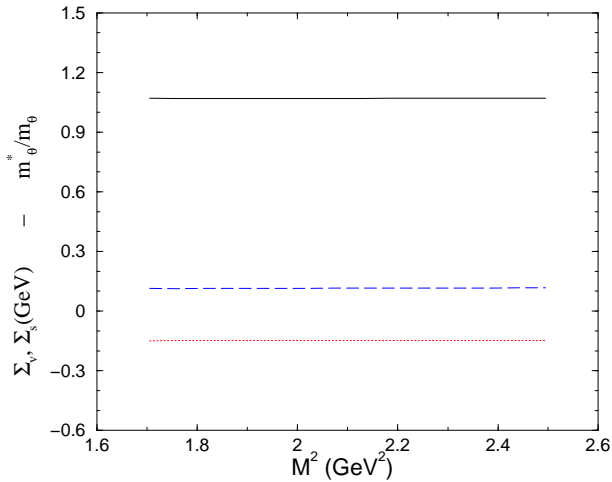,width=8cm,angle=0}}
\caption{The ratio $m_\Theta^*/m_\Theta$ (solid line), the vector and 
scalar self-energies (dotted and dashed lines respectively) as functions of
the Borel parameter $M^2$, for $y=0.3$.}
\end{figure}

The optimized results  for the ratio $\mtes/\mte$ (solid line) and for the
scalar (dashed line) and vector (dotted line) self-energies as functions of 
$M^2$ are plotted in Fig.~2 for $y=0.3$. We see that the curves are quite flat,
indicating a weak dependence of the predicted results  on $M^2$. The results
are also not sensitive to the value of $\veq$. Using $\veq=0$ instead
of $\veq=270\MeV$ does not alter our results. On the other hand, the results 
are very sensitive to
the value of $y$. In the range $0\leq y\leq0.5$ we got 
$50\MeV\leq\Sigma_s\leq150\MeV$ and $-90\MeV\geq\Sigma_v\geq-190\MeV$. What is
very interesting is that the sum $\Sigma_s+\Sigma_v$, which can be associated 
with the depth of the $\Theta^+$ potential in nuclear matter, remains 
independent of $y$ and is about $-40\MeV$. However, this value
is very sensitive to the value of the gluon condensate. If we change the value
of the gluon condensate to $\me{\gluoncon}=0.24~\GeV^4$, as used in 
ref.~\cite{iof}, we get (for $y=0.3$): $\Sigma_s\sim90\MeV$ and
$\Sigma_v\sim-180\MeV$ which would imply in a potential depth of about
$-90\MeV$.

The depth of the
$\Theta^+$ potential in nuclear matter, $U$, was studied in ref.~\cite{zho} 
using a relativistic mean field framework, where $\Theta^+$ couples with 
scalar and isoscalar-vector mesons. They found $-90\MeV\leq U\leq-45\MeV$
depending on the values used for $m_N^*/m_N$ (the ratio of the nucleon mass),
and the coupling constants. In our work we found that
the potencial depth depends strongly on the value of the gluon condensate.
In the range $0.47\GeV^4\leq\me{\gluoncon}\leq0.24~\GeV^4$ we got
$-40\MeV\geq U\geq-90\MeV$, which is compatible with the findings of 
ref.\cite{zho}.

It is also interesting that, in
spite of $\mtes$ being bigger than $\mte$, the energy of the quasi-$\Theta$
in nuclear matter, $E_q$, is smaller than $\mte$ and it is the most stable
result of our calculation. It does not depend on $y$ neither on
 $\me{\gluoncon}$. While $\mte$, $\mtes$ and $\Sigma_v$ varies significantly
with $y$ and $\me{\gluoncon}$ we got $E_q\sim1.60\GeV$ for all the values
of the parameters (the smallest value obtained for the $\Theta^+$ mass
was $\mte\simeq1.64\GeV$).

\begin{figure}[h] \label{fig3}
\centerline{\psfig{figure=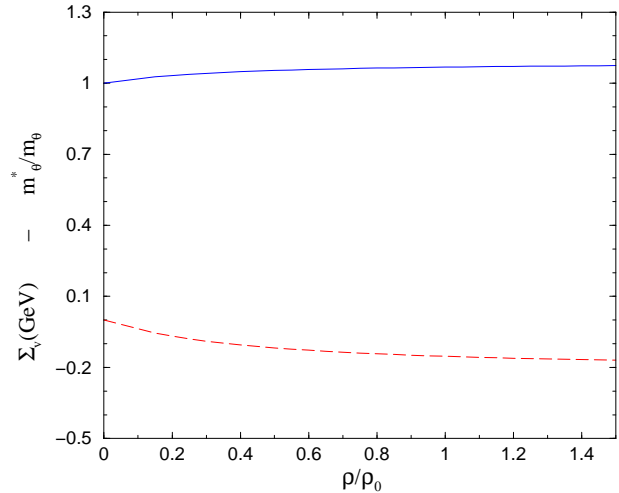,width=8cm,angle=0}}
\caption{The ratio $\mtes/\mte$ (solid line) and the vector self-energy 
(dashed line) as  functions of the baryon density, for $y=0.3$.}
\end{figure}

All the results given above were obtained at the nuclear matter saturation 
density
$\rho_0$. In Fig.~3 we show the density dependence of the scalar and vector 
self-energies. We see that most of the variation happens for 
$\rho_N<0.5\rho_0$.
Since we are expanding the density dependence of the condensates up to first
order in the nuclear density, the obtained density dependence of our results
is compatible with the approximations made.

We would like to point out that, since our sum rules do not depend on the 
four-quark condensates, our results are free from the uncertainties associated 
with the density dependence of these condensates, which were the biggest 
source of uncertainty in the case of the nucleon and hyperons studied in 
refs.~\cite{srnm1,srnm2,srnm3}. In that case, the scalar self-energy was very 
sensitive to the density dependence of the four-quark condensates, and
one could even get a repulsive net self-energy. The results were a bit 
more stable for the vector self-energy. Comparing the values of the vector 
self-energies one has from refs.~\cite{srnm1,srnm2,srnm3}: 
$\Sigma_v^\Lambda/\Sigma_v^N\simeq0.3-0.4$, $\Sigma_v^\Sigma/\Sigma_v^N
\simeq0.8-1.1$. And here we got $|\Sigma_v^\Theta|/\Sigma_v^N\simeq0.3-0.5$.
In terms of relativistic hadronic model these results would imply that
the coupling of the hyperon $\Sigma$ to the Lorentz vector field is very 
similar to the corresponding nucleon coupling, while the coupling of the
hyperon $\Lambda$ is similar to the coupling of $\Theta$ and both are much
weaker than the correponding nucleon coupling. These results can be understood
in terms of the interpolating fields used to study these baryons. For
$\Lambda$ there is a $[ud]$ diquark with spin and isospin zero plus the 
strange quark, that carries the spin of $\Lambda$. Assuming no admixture of
strange quark content in the vector meson $\omega$ (the Lorentz vector field),
there will be no coupling between $\Lambda$ and $\omega$. The same happens
in the case of $\Theta^+$ (see Eq.~(\ref{eta})). The situation is different 
in the case of $\Sigma$, since in its interpolating field there is a
$[qs]$ diquark with spin zero. Therefore, it is the light quark that carries 
the spin of $\Sigma$, as in the nucleon case.

As a final remark, we would like to mention that in ref.~\cite{mil} it was 
suggested that if the parity of $\Theta^+$ were
positive, then there would be a strong attractive $\Theta-$nucleon 
interaction. In our calculation we got a negative parity for $\Theta^+$. Its 
interaction with the nucleon is still attractive, and roughly of the same 
order of the nucleon-nucleon interaction. Bigger attraction is obtaned with
smaller values of the gluon condensate.

\vspace{1cm}
 
\underline{Acknowledgements}: 
This work has been supported by CNPq and  FAPESP (Brazil). We thank G. Krein 
for fruitful discussions. 


\end{document}